\documentclass[aip,graphicx]{revtex4-1}

\usepackage{graphicx}
\usepackage{bm}
\usepackage{siunitx}
\usepackage[T1]{fontenc}

\draft 

\begin{document}


\title{Mid-Infrared Photon-Pair Generation in AgGaS$_2$} 



\author{Mohit Kumar}
\email[]{mohit.kumar@uni-jena.de}
\affiliation{Institute of Applied Physics, Abbe Center of Photonics, Friedrich Schiller University Jena, Albert-Einstein-Str. 15, 07745 Jena,
Germany}
\author{Pawan Kumar}
\affiliation{Institute of Applied Physics, Abbe Center of Photonics, Friedrich Schiller University Jena, Albert-Einstein-Str. 15, 07745 Jena,
Germany}

\author{Andres Vega}
\affiliation{Institute of Applied Physics, Abbe Center of Photonics, Friedrich Schiller University Jena, Albert-Einstein-Str. 15, 07745 Jena,
Germany}

\author{Maximilian A. Weissflog}
\affiliation{Institute of Applied Physics, Abbe Center of Photonics, Friedrich Schiller University Jena, Albert-Einstein-Str. 15, 07745 Jena,
Germany}
\affiliation{Max Planck School of Photonics, Albert-Einstein-Str. 6, 07745 Jena, Germany}

\author{Thomas Pertsch}
\affiliation{Institute of Applied Physics, Abbe Center of Photonics, Friedrich Schiller University Jena, Albert-Einstein-Str. 15, 07745 Jena,
Germany}
\affiliation{Max Planck School of Photonics, Albert-Einstein-Str. 6, 07745 Jena, Germany}

\affiliation{Fraunhofer Institute for Applied Optics and Precision Engineering, Albert-Einstein-Str. 7, 07745 Jena, Germany}

\author{Frank Setzpfandt}
\affiliation{Institute of Applied Physics, Abbe Center of Photonics, Friedrich Schiller University Jena, Albert-Einstein-Str. 15, 07745 Jena,
Germany}


\date{\today}

\begin{abstract}
We demonstrate non-degenerate photon-pair generation by spontaneous parametric down conversion in a silver gallium sulfide AgGaS$_2$ crystal. By tuning the pump wavelength, we achieve phase matching over a large spectral range. This allows to generate idler photons in the mid-infrared spectral range above \SI{6}{\micro\metre} wavelength with corresponding signal photons in the visible. Also, we show photon pair generation with broad spectral bandwidth. These results are a valuable step towards the development of quantum imaging and sensing techniques in the mid-infrared.
\end{abstract}


\maketitle 


Quantum-correlated photon pairs with non-degenerate wavelengths are the key ingredient in many quantum optics applications\cite{GilaberteBasset2019,Mukamel2020}. This is especially true for applications based on nonlinear interferometers, where one photon is detected, whereas only its correlated partner photon probes a test object in a different wavelength range\cite{Lemos2014,Kalashnikov2016,Chekhova2016}. Thus, a photon-pair source providing pairs with one photon in the visible and one in the infrared, used in a nonlinear interferometer, provides a powerful tool to perform measurements in the infrared spectral range without using an infrared detector, as the object information can be obtained using highly efficient and well-developed visible range detectors like charge-coupled device (CCD) and complementary metal–oxide–semiconductor (CMOS) sensors. Using this technique, many applications have been demonstrated in quantum imaging\cite{Lemos2014}, microscopy\cite{Kviatkovskyeabd0264}, spectroscopy\cite{Kalashnikov2016,Lindner2020,Lindner2021}, and optical coherence tomography\cite{Paterova_2018} in recent years. On the other hand, heralded single-photon sources can be advantageous for many quantum-optics applications, especially for quantum ghost spectroscopy\cite{GilaberteBasset2019,PhysRevA.69.013806}. This is because a heralded single-photon source enables to perform measurements with a better signal-to-noise ratio (SNR) in comparison to using a classical light source, demonstrated by Kalashnikov and co-workers\cite{PhysRevX.4.011049}.

However, existing photon-pair sources limit such applications to wavelengths below \SI{5}{\micro\metre} in the mid-infrared (MIR) spectral range. The reason is, that for most typically used nonlinear crystals such as KTiOPO$_4$ (KTP) and LiNbO$_3$ (LN), photons at longer wavelengths get absorbed by the nonlinear medium itself, which prohibits efficient photon-pair generation. On the other hand, nonlinear crystals such as ZnGeP$_2$ (ZGP) and AgGaSe$_2$ (AGSe), which are transparent in the long-wavelength region, suffer from stronger absorption in the visible and near infrared spectral range\cite{McCracken2018,Maidment2018,Isaenko_2016}. Nevertheless, a highly non-degenerate and widely tunable correlated photon-pair source with signal photons in the visible and idler photons in the MIR spectral range is essential to access the technologically important mid-IR spectral range which covers the unique fingerprints of many organic and inorganic materials as well as the atmospheric transmission window (8-\SI{13}{\micro\metre}).

In this work, we demonstrate photon-pair generation in silver gallium sulfide AgGaS$_2$ (AGS), a nonlinear crystal with a transparency range spanning from \SI{0.5}{\micro\metre} in the visible to \SI{13}{\micro\metre}\cite{David2005} in the MIR. We generate idler photons in the MIR above \SI{6}{\micro\metre} wavelength with corresponding signal photons in the visible. We also demonstrate the tunability of the source by changing the pump wavelength to tune the signal-idler central wavelengths. Furthermore, we characterize the generated pairs using correlation measurements and demonstrate a high generation rate and purity.

We generate photon pairs by spontaneous parametric down conversion (SPDC), where pump photons of higher energy spontaneously split into pairs of signal and idler photons due to a second-order nonlinearity. The signal and idler photons have to fulfil energy and momentum conservation, which induces spectral and spatial correlations between the two photons of a pair. The efficiency of the SPDC process is primarily controlled by the phase-matching condition, with the maximum efficiency for zero phase mismatch $\Delta k(\lambda_s, \lambda_i)=\left|\mathbf{k}(\lambda_p)-\mathbf{k}(\lambda_s)-\mathbf{k}(\lambda_i)\right|=0$, where $\mathbf{k}$ is the wavevector with $\left|\mathbf{k}(\lambda_x)\right| = 2\pi n(\lambda_x)/\lambda_x$, $\lambda_x$ is the wavelength, $n(\lambda_x$) is the wavelength-dependent refractive index, and $x = (p, s, i)$ represents the pump, signal and idler, respectively. In the plane wave approximation, for a phase mismatch $\Delta k$ and crystal length $L$, the joint spectral probability (JSP) of the generated signal / idler photon pair is given by $I (\lambda_s, \lambda_i)\propto \text{sinc}^2(\Delta k (\lambda_s, \lambda_i)L/2)$\cite{Boyd2006}.

In AGS, phase matching can be achieved over a large spectral range in the SPDC type-I configuration\cite{Boyd1971}, where the pump photon is polarized along the extra-ordinary (e) and the signal and idler photons are polarized along the ordinary (o) axis of AGS crystal. In Fig. 1 we sketch this SPDC configuration. The phase mismatch is\cite{Boyd2006,Boyd1971,Simon2017}
\[\Delta k(\lambda_s, \theta_s)= k^e_p(\lambda_p,\theta_{PM})-k^o_s(\lambda_s)\mathrm{cos}\theta_s-k^o_i(\lambda_i)\mathrm{cos}\theta_i,\]
where $\theta_{PM}$ is the phase matching angle, $\theta_s$ and $\theta_i$ are the emission angles of signal and idler photons inside the crystal with respect to the pump wave vector. We characterize the photon pairs generated in AGS using two methods sketched in the insets of Fig. 1, classical spectroscopy to measure the wavelengths of generated signal photons and a fiber-based Hanbury Brown - Twiss (HBT) setup to measure the correlation between the generated photon pairs.

\begin{figure}
\includegraphics{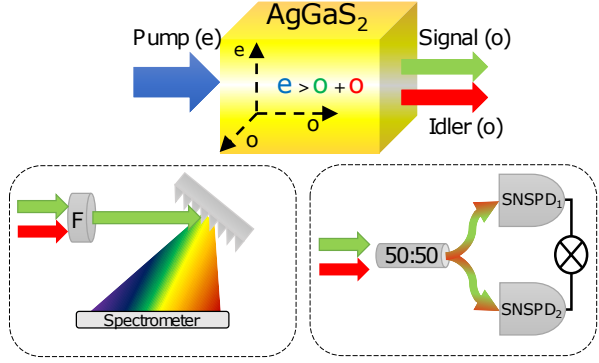}
\caption{\label{fig} Schematic of SPDC type-I configuration in AGS crystal, and two characterization schemes. Left inset: classical spectroscopy for the measurement of generated signal wavelengths, right inset: Fiber based Hanbury Brown - Twiss (HBT) setup for correlation measurement between the generated photon pairs. SNSPD are superconducting nanowire single-photon detectors}
\end{figure}

For our experiments, we use a \SI{2}{\milli\metre} long AGS crystal with a phase matching angle $\theta_{PM}=90^{\circ}$ (ASCUT, Berlin). Phase matching is also possible for other angles, but we deliberately choose a $90^{\circ}$ angle to have the degenerate as well as shorter signal wavelengths of photon pairs within the spectral sensing range of our detectors\cite{Boyd1971}.
As pump sources, we use a tuneable Ti:Sapphire CW laser (M2 SolsTiS, tuning range $\lambda$ = \SI{0.7}{\micro\metre} to \SI{1.0}{\micro\metre}) and a He:Ne laser (Thorlabs, \SI{0.63}{\micro\metre}). A lens of \SI{40}{\milli\metre} focal length is used to create an approximate pump beam diameter of \SI{35}{\micro\metre} at the AGS crystal. We use a collection optics with numerical aperture (NA) of 0.012 to collect the collinearly generated photon pairs within an emission angle $\pm 1^{\circ}$. The photon pairs are then separated from the pump using suitable long pass filters (Thorlabs). We perform all the measurements at room temperature (\SI{22}{\celsius}).

First, we perform classical spectroscopy of the signal photon, as sketched in the left inset of Fig. 1, using an InGaAs-detector-based spectrometer sensitive from \SI{0.7}{\micro\metre} to \SI{1.65}{\micro\metre}, which cannot detect the idler photons at longer wavelengths. We pump the AGS crystal with different pump wavelengths starting from \SI{0.88}{\micro\metre} to \SI{0.63}{\micro\metre} (minimum available pump wavelength) and measure corresponding generated signal-photon spectra. The experimentally measured normalized spectra for the pump wavelengths \SI{0.88}{\micro\metre}, \SI{0.83}{\micro\metre} and \SI{0.63}{\micro\metre} are shown in Fig.~2(a), exhibiting maxima for signal wavelengths at \SI{1.55}{\micro\metre}, \SI{1.17}{\micro\metre} and \SI{0.70}{\micro\metre} respectively. Using the phase-matching conditions in the SPDC type I configuration and the Sellmeier equations of the AGS crystal\cite{Takaoka1999}, we calculate the expected signal photon spectra for the same pump wavelengths as in the experiment. The experimentally measured spectra (solid lines) closely agree with the calculated signal photon spectra (dashed lines).

To further characterize the spectra of generated signal photons, we perform a measurement to show that AGS is generating photon pairs with broad bandwidth, which mainly arises due to non-collinearly phase-matched photon pairs in the SPDC type-I configuration. For this measurement, we use a collection optics with increased NA of 0.15, enabling to collect signal photons within $\pm 8.5^{\circ}$ emission angle. The measured spectrum (solid green line) for \SI{0.83}{\micro\metre} pump wavelength is shown in Fig. 2(a). We observe a ($1/e^2$)-bandwidth of $\sim $\SI{0.20}{\micro\metre} for the signal wavelengths, which is broad compared to the measured bandwidth of $\sim $\SI{25}{\nano\metre} within the emission angle of $\pm1^{\circ}$ (solid red lines) for the same pump wavelength. Based on the observed bandwidth of signal photons and phase matching condition, the correlated idler wavelengths can be expected to have a broad bandwidth of $\sim$\SI{2.50}{\micro\metre}, spanning from \SI{2.77}{\micro\metre} to \SI{5.27}{\micro\metre}. Also here, the experimentally measured broadband spectra (solid green lines) are in a good agreement with the calculated spectra (dashed green lines). By tuning pump wavelength, photon pairs with a broad spectral bandwidth can also be generated in other spectral ranges (within the transparency range of AGS crystal). In addition, one can select the appropriate bandwidth by controlling the collection angle. This broadband feature of generated photon pairs along with the spectral tunability makes AGS crystals useful in increasing the spectral coverage of quantum imaging\cite{Kviatkovskyeabd0264} and sensing\cite{Lindner2020} applications across the MIR range.

\begin{figure}
\includegraphics{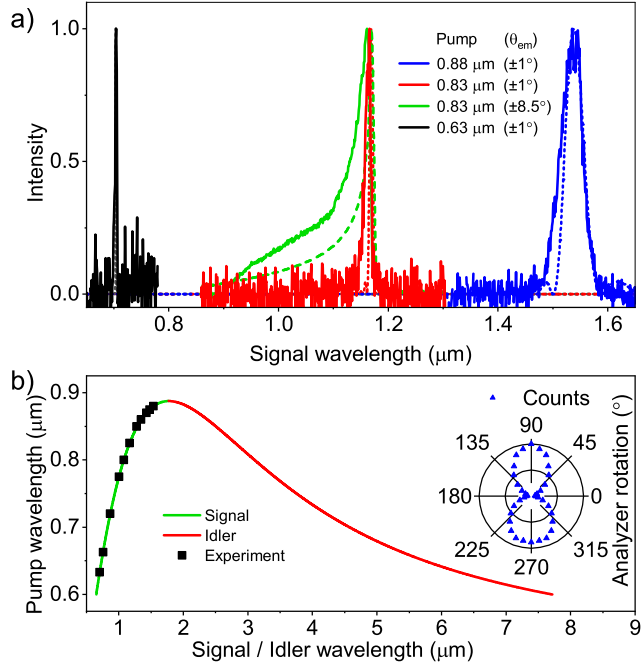}
\caption{\label{fig} (a) Experimentally observed (solid lines) and simulated (doted lines) spectra of the signal photons for pump wavelengths of \SI{0.88}{\micro\metre}, \SI{0.83}{\micro\metre} and \SI{0.63}{\micro\metre}, corresponding to their emission angles ($\theta_{em}$). (b) The calculated SPDC type-I phase-matching curve of AGS crystal as a function of pump wavelength, signal wavelength (green lines), and idler wavelength (red lines) as well as experimentally measured wavelengths (Black squares). Inset: Polarization dependent measurements of SPDC Type-I generated signal photons, where $0^{\circ}$ analyser position is parallel to pump polarization.}
\end{figure}

In Fig. 2(b), we plot the calculated phase-matched signal (green) and idler (red) wavelengths as a function of pump wavelengths, for a fixed phase matching angle of  $\theta_{PM} = 90^{\circ}.$ We find phase matching for wavelength-degenerate SPDC at a pump wavelength of \SI{0.89}{\micro\metre}, corresponding to signal and idler wavelengths of \SI{1.77}{\micro\metre} in the near-infrared, and for non-degenerate SPDC at pump wavelengths shorter than \SI{0.89}{\micro\metre}. We observe that by reducing the pump wavelength we can increase the spectral separation of non-degenerate photon pairs, and can generate highly non-degenerate photon pairs over a large spectral range. In addition, we clearly see that for shorter pump wavelengths, the wavelengths of the strongly non-degenerate photon pairs are in the visible range for signal photons and in the MIR range for idler photons. For all experimentally tested pump wavelengths from \SI{0.88}{\micro\metre} to \SI{0.63}{\micro\metre}, we plot the wavelengths of the observed spectral maxima (black squares) in Fig.~2(b), showing a good agreement with the calculated phase matched signal wavelengths (green line). This clearly demonstrates generation of signal wavelengths in a broad spectral range from \SI{1.55}{\micro\metre} to \SI{0.70}{\micro\metre}. The correlated idler wavelengths are expected to be spanning from \SI{2.0}{\micro\metre} to \SI{6.3}{\micro\metre}. 

Based on this spectroscopic analysis of the signal photons and the phase-matching calculations, we conclude that the obtained idler wavelength for a pump of \SI{0.63}{\micro\metre} was at \SI{6.3}{\micro\metre}, corresponding to the measured signal wavelength of \SI{0.70}{\micro\metre}. By using pump wavelengths lower than \SI{0.63}{\micro\metre}, idler photon generation can be tuned above \SI{6.3}{\micro\metre} in the mid-IR spectral range, while signal photon remains in the visible spectral range. This feature makes the AGS photon-pair source suitable for mid-IR quantum imaging and spectroscopy applications using only visible light detectors and sources.

We verify the polarization of photon pairs generated through type-I SPDC by placing an analyzer before the spectrometer. For a pump wavelength of \SI{0.88}{\micro\metre} and fixed input pump polarization (parallel to the extra-ordinary axis of the crystal), we rotate the analyzer and measure spectra of the generated signal photons in $10^{\circ}$ steps of analyzer rotation. In the inset of Fig. 2(b), we plot the measured counts corresponding to spectral maxima as a function of analyzer rotation. We observe almost zero counts when the analyzer is parallel to the pump polarization and maximum counts with an analyzer rotation of $90^{\circ}$. This clearly shows that the generated signal photons are orthogonally polarized to the pump photons, as expected in SPDC type-I configuration. Here, correlated idler photons are expected to be generated co-polarized with signal photons as all the experimentally measured signal wavelengths satisfied the type-I phase matching condition.

To demonstrate the generation of correlated photon pairs we use an HBT setup as sketched in the right inset of Fig. 1(a). This consists of a 50:50 fiber beam splitter, two superconducting nanowire single-photon detectors (SNSPD – Single Quantum EOS, timing jitter $\leq$ \SI{25}{\pico\second}) and a correlation electronics (IDQuantique ID800, temporal resolution \SI{81}{\pico\second}) that measures the difference in the arrival times of the two photons. We pump the AGS crystal with a pump power of \SI{1.5}{\milli\watt} at the wavelength of \SI{0.89}{\micro\metre}, where degenerate photon pairs with $\lambda_s$ = $\lambda_i$ = \SI{1.77}{\micro\metre} are generated. The coincidence counts as a function of the time difference between the arrival times of the photons are plotted in Fig. 3(a). We observe a clear coincidence peak (red) at zero time difference with a coincidence count rate of $\sim$\SI{5.5}{\kilo\hertz} and a peak width of \SI{0.65}{\nano\second}. The observed peak demonstrates generation of photon pairs. A similar measurement was performed with a slightly shorter pump wavelength of \SI{0.88}{\micro\metre}, where non-degenerate photon pairs with $\lambda_s$ = \SI{1.68}{\micro\metre} and $\lambda_i$ = \SI{1.87}{\micro\metre} are generated. Also, here we observed a strong coincidence peak (black) with a coincidence count rate of $\sim$\SI{6.4}{\kilo\hertz}, plotted in Fig. 3(a). These measured coincidence peaks confirm the generation of photon pairs with degenerate as well as non-degenerate wavelengths. The direct observation of photon pairs with even longer idler wavelength was not possible due to limitations in the spectral sensitivity of our detection system.

\begin{figure}[h]
\includegraphics{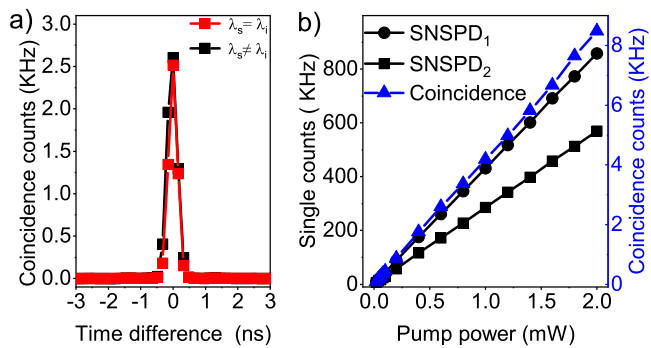}
\caption{\label{fig} (a) Measured coincidence counts as a function of arrival times difference of the two photons, for degenerate photon pairs (red) where $\lambda_s$ = $\lambda_i$ = \SI{1.77}{\micro\metre} and non-degenerate photon pairs (black) where $\lambda_s$ = \SI{1.68}{\micro\metre} and $\lambda_i$ = \SI{1.87}{\micro\metre}, with a clear coincidence peak at zero-time difference. (b) Measured single counts on the two SNSPDs and coincidence count rate as a function of pump power.}
\end{figure}

We further investigate the properties of the AGS SPDC source by measuring single, coincidence and accidental count rates at different pump powers. In Fig. 3(b), we plot the single count rates $N_1$ and $N_2$ on the two SNSPD$_s$ with subtracted dark counts, and the true coincidence count rate $N_{12}$ as a function of pump power, where $N_{12}=N_{12, raw}-N_{acc}$. The raw coincidence counts $N_{12,raw}$ are the sum of all measured counts under the coincidence peak and the accidental counts $N_{acc}$ are the sum of all counts outside the coincidence peak for the same peak width of \SI{0.65}{\nano\second}. We observe linearly increasing single and coincidence count rates with increasing pump power. This shows that we are still in the spontaneous regime of pair generation. The observed true  coincidence count rate is $N_{12}$ = \SI{4.2}{\kilo\hertz\per\milli\watt}. Based on the measured single and coincidence counts, we estimate the photon pair generation rate $N_{pair}$ = \SI{14.6}{\mega\hertz\per\milli\watt} using $N_{pair} = N_1N_2 / 2N_{12}$\cite{Simon2017}, and a conversion efficiency of 3.3 x$10^{-9}$, which considers the limited collection and detection efficiencies of our detection system. In this calculation, we have also considered a 50$\%$ loss in the coincidence counts, as the random splitting of the photon pairs by 50:50 fiber beam splitter leads to a 50$\%$ chance that photon pairs go to the same detector.

\begin{figure}[h]
\includegraphics{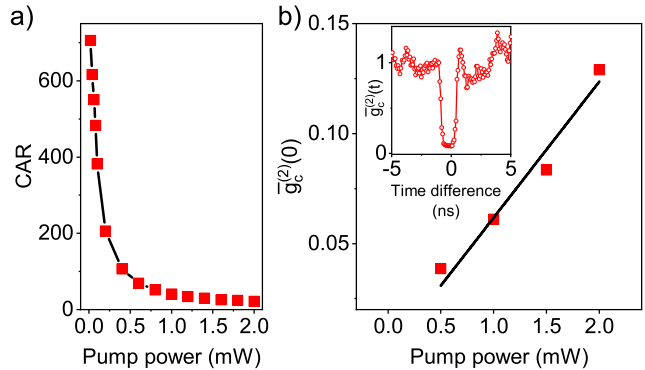}
\caption{\label{fig} (a) Measured value of coincidence to accidental ratio (CAR) as a function of pump power. (b) Time-averaged conditioned second-order correlation function $\bar{g}^{(2)}_c(0)$ at the zero-time difference as a function of pump power. Inset: For a pump power of \SI{1}{\milli\watt} the measured and normalized value of $\bar{g}^{(2)}_c(t)$ with a clear heralded antibunching dip at zero-time difference.}
\end{figure}

In Fig. 4(a) we plot the coincidence-to-accidental ratio (CAR) as a function of pump power, where CAR $ = {N_{12}}/{N_{acc}}$. For a pump power of \SI{20}{\micro\watt}, we observe a CAR of 706 corresponding to a pair generation rate of \SI{318}{\kilo\hertz}. For a pump power of \SI{2}{\milli\watt}, we observe a CAR value of 22 corresponding to a pair generation rate of \SI{28.8}{\mega\hertz}. The observed decrease in the CAR value with increasing pump power indicates a higher probability of multiphoton events which contribute to the accidental counts.

To characterize the contribution of unwanted multi-photon states, and to demonstrate the possibility of generating heralded single photons in our SPDC source, we measure the second-order correlation function using a conditioned HBT setup. For this measurement, we modify one arm of the HBT setup used for the correlation measurements, by adding a 50:50 beam splitter and a third single-photon detector. In this configuration, when one photon of the pair is registered on SNSPD$_1$, it heralds the presence of a correlated partner photon in the system. Conditionally the partner photon can only be registered either on  SNSPD$_2$ or  SNSPD$_3$. Therefore, in the case of only photon-pair generation, there will be zero simultaneous event registration on all three  SNSPD$_s$. As a result, we expect a reduced value of the conditioned second-order correlation function $\bar{g}^{(2)}_c(t)$ at the zero-time difference. The normalized time-averaged $\bar{g}^{(2)}_c(t)$ is\cite{Bocquillon2009,Bettelli2010,Hckel2011}
\begin{center}
$\bar{g}^{(2)}_c(t)=\frac{N_{1}N_{123}}{N_{12}N_{13}}$,    
\end{center}
where $N_{123}$ are the 3-fold coincidence between all three SNSPD$_s$, $N_{12}$ and $N_{13}$ are 2-fold coincidence counts as measured in the previous section, and $N_{1}$ are the single counts registered on SNSPD$_1$. For an ideal photon-pair source the expected $\bar{g}^{(2)}_c(0)=0$. In our experiment, first we record the arrival times of the photons (time tags) at all SNSPD$_s$ using the time-tagging device with a temporal resolution of \SI{81}{\pico\second}. Then we process these time tags to extract the $\bar{g}^{(2)}_c(t)$. For a pump power of \SI{1}{\milli\watt} and a pump wavelength of \SI{0.89}{\micro\metre}, we plot the obtained value of $\bar{g}^{(2)}_c(t)$ as a function of the arrival time difference (t) between SNSPD$_2$ and SNSPD$_3$ in the inset of Fig. 4(b). Here, we clearly observe the expected signature with an obtained value of $\bar{g}^{(2)}_c(0)=0.061$, at an estimated pair generation rate of \SI{14.6}{\mega\hertz}. We perform similar measurements at various pump powers and plot the measured $\bar{g}^{(2)}_c(0)$ as a function of pump power in Fig. 4(b). We observe an increase in $\bar{g}^{(2)}_c(0)$, as the multi-photon generation probability increases with the pump power, increasing the likelihood of three-fold coincidence events. A small value of $\bar{g}^{(2)}_c(0)=0.038$ is obtained for a pump power of \SI{0.5}{\milli\watt}. At the cost of pair generation rate, one can further reduce $\bar{g}^{(2)}_c(0)$ by reducing the pump power. These measurements confirm that the AGS crystal can generate photon pairs with good overall count rates while still operating in the spontaneous pair generation regime at few mW of pump power.

To summarize, we have demonstrated AGS as a broadly tuneable correlated photon-pair source generating non-degenerate photon pairs with idler wavelengths well in the MIR corresponding to signal wavelengths in the visible. We have shown spectral bandwidth dependencies of photon pairs on the emission angle. In addition, we have demonstrated AGS as a potential source of heralded single photons. The spectral range of photon pairs can be further extended up to \SI{13}{\micro\metre} by using pump wavelengths lower than \SI{0.63}{\micro\metre}. Alternatively, this could also be achieved for a fixed pump wavelength, either by heating (temperature tuning) or rotating (angular tuning) the AGS crystal. Our results are of importance for the development of quantum imaging and sensing techniques in the mid-infrared.

\begin{acknowledgments}
We thank E. Santos and V. Gili for their insightful comments.
This work was supported by the Thuringian Ministry for Economy,
Science, and Digital Society and the European Social Funds (2017 FGR 0067); European Union's Horizon 2020 research and innovation programme under grant agreement No 899580; the German Federal Ministry of Education
and Research (Nos. FKZ 13N14877); and the
Deutsche Forschungsgemeinschaft (DFG, German Research
Foundation, Project ID 327470002, 407070005, and ID 398816777-SFB 1375).

\end{acknowledgments}

\section*{data availability}
The data that support the findings of this study are available from the corresponding author upon reasonable request.

\bibliography{AGS_MIR_PPS.bib}

\end{document}